\documentclass[journal]{IEEEtran}
\usepackage[latin9]{inputenc}
\usepackage{xcolor}
\usepackage{pdfcolmk}
\usepackage{amsmath}
\usepackage{amssymb}
\usepackage{graphicx}
\PassOptionsToPackage{normalem}{ulem}
\usepackage{ulem}
\usepackage{fancyhdr}
 
\makeatletter

\providecolor{lyxadded}{rgb}{0,0,1}
\providecolor{lyxdeleted}{rgb}{1,0,0}

\DeclareRobustCommand{\lyxsout}[1]{\ifx\\#1\else\sout{#1}\fi}

\IEEEoverridecommandlockouts
\usepackage{cite}
\usepackage{amsfonts}\usepackage{algorithmic}
\usepackage{subfigure}
\usepackage{textcomp}
\usepackage{xcolor}
\def\BibTeX{{\rm B\kern-.05em{\sc i\kern-.025em b}\kern-.08em
    T\kern-.1667em\lower.7ex\hbox{E}\kern-.125emX}}

\makeatother

\begin{document}

\title{High Power Backward Wave Oscillator using Folded Waveguide with Distributed
Power Extraction Operating at an Exceptional Point}

\author{\IEEEauthorblockN{Tarek Mealy, Ahmed F. Abdelshafy and Filippo Capolino}
\IEEEauthorblockA{\textit{Department of Electrical Engineering and Computer Science,
University of California, Irvine, CA 92697 USA} \\
tmealy@uci.edu, abdelsha@uci.edu and f.capolino@uci.edu}}
\maketitle
\thispagestyle{fancy}
\begin{abstract}
The concept of exceptional point of degeneracy (EPD) is used to conceive
a degenerate synchronization regime that is able to enhance the level
of output power and power conversion efficiency for backward wave
oscillators (BWOs) operating at millimeter-wave and Terahertz frequencies.
Standard BWOs operating at such high frequency ranges typically generate
output power not exceeding tens of watts with very poor power conversion
efficiency in the order of 1\%. The novel concept of degenerate synchronization
for the BWO based on a folded waveguide is implemented by engineering
distributed gain and power extraction along the slow-wave waveguide.
The distributed power extraction along the folded waveguide is useful
to satisfy the necessary conditions to have an EPD at the synchronization
point. Particle-in-cell (PIC) simulation results shows that BWO operating
at an EPD regime is capable of generating output power exceeding 3
kwatts with conversion efficiency of exceeding 20\% at frequency of
88.5 GHz.
\end{abstract}

\begin{IEEEkeywords}
Exceptional point of degeneracy, Degenerate synchronization, Slow-wave
structures, Backward-wave oscillators, High power microwave.
\end{IEEEkeywords}

\section{Introduction}

The capability to generate significant power at millimeter-wave and
terahertz (THz) frequencies using vacuum electronics sources has motivated
many investigations due to the their high demand on applications such
as imaging, spectroscopy and communications \cite{booske2008plasma,booske2011vacuum,armstrong2012truth,joye2014demonstration,armstrong2018compact,9262150}.
Vacuum electronic devices operating at millimeter-wave and THz frequencies
often use folded waveguides (serpentine-shaped waveguide) as shown
in Fig. \ref{Fig:SWS_No_DPE}. The advancement of fabrication technologies
such as LIGA (LIthographie, Galvanoformung, Abformung) have made it
easy to fabricate and develop vacuum electronic devices operating
at these high frequencies \cite{shin2006experimental,sengele2009microfabrication,dobbs2010design,feng2011study,joye2013microfabrication,joye2014demonstration}.

Exceptional points of degeneracy (EPD) are points in parameter space
of a system at which two or more eigenmodes coalesce. Despite most
of the published work on EPDs are related to parity time (PT) symmetry
\cite{bender1998real,klaiman2008visualization}, the occurrence of
EPDs does not necessarily require a system to exactly satisfy the
PT symmetry condition. In general, but not always \cite{othman2017experimental,abdelshafy2018exceptional,mealy2020general}
, the occurrence of EPD in waveguides requires simultaneous presence
of gain and loss \cite{othman2017theory,abdelshafy2018exceptional}.
Instead of using losses, the EPD in this work is enabled by the distributed
power extraction (DPE) from the folded waveguide as shown in Fig.
\ref{Fig:SWS_DPE}. The ideal concept of simultaneous radiation losses
and distributed gain in two coupled waveguides leading to an EPD was
already discussed in \cite{othman2017theory} in a theoretical setting.
In this paper the concept is achieved using a single serpentine waveguide
coupled to an electron beam (e-beam). The energy extracted from the
e-beam and delivered to the guided electromagnetic (EM) mode is considered
as a distributed gain from the waveguide perspective, whereas DPE
represents extraction ``losses'' and not mere dissipation \cite{mealy2019backward,mealy2019exceptional,mealy2020exceptional}.
The distributed extracted power from the discrete waveguide ports
along the serpentine (Fig. \ref{Fig:BWOs_All_SER}) could be directed
toward an array antenna and hence radiated generating a collimated
EM beam or could be collected in an external waveguide after proper
optimization for power combining.

\textcolor{black}{In \cite{mealy2019exceptional}, we have studied
the theoretical and idealistic analysis of a BWO with degenerate synchronization
operating at an EPD (we named it EPD-BWO) using a generalized Pierce
model \cite{pierce1951waves} that accounts for waveguide with distributed
power extraction modeled as losses.} Here we show analytically and
numerically using particle-in-cell (PIC) simulations that an EPD-BWO
is characterized by the asymptotic trend of the \textit{starting}
e-beam dc current that decreases quadratically with slow wave structure
(SWS) length to a non-vanishing value\textcolor{black}{{} that can be
properly designed based on the required output power. In this paper,
we focus on the realization and implementation of the degenerate synchronization
of the EPD-BWO at millimeter-waves since it is very challenging to
generate significant power levels at such high frequencies. Though
we do not show it here, the concept of degenerate synchronization
is expected to be advantageous also at THz frequencies. The use of
EPD enables to have higher starting current for oscillation which
indicates higher level of power extraction from the e-beam kinetic
energy.}

Works on BWOs operating at millimeter-wave and THz have reported a
generated output power not exceeding tens of watts with power conversion
efficiency around 1\% \cite{choi1996experimental,bhattacharjee2004folded,feng2011study,nguyen2014design,cai2016study,baig2017performance,shu2018demonstration,9262150}.
In this paper we employ the concept of EPD to enhance such poor efficiency
and to increase the level of output power. We asses the advancements
in the performance of BWO operating at an EPD (EPD-BWO) over a standard
BWO (STD-BWO) using particle-in-cell (PIC) simulations.

\section{Implementation of degenerate synchronism regime based on EPD in a
folded waveguide }

\begin{figure}
\begin{centering}
\subfigure[]{\includegraphics[width=1\columnwidth]{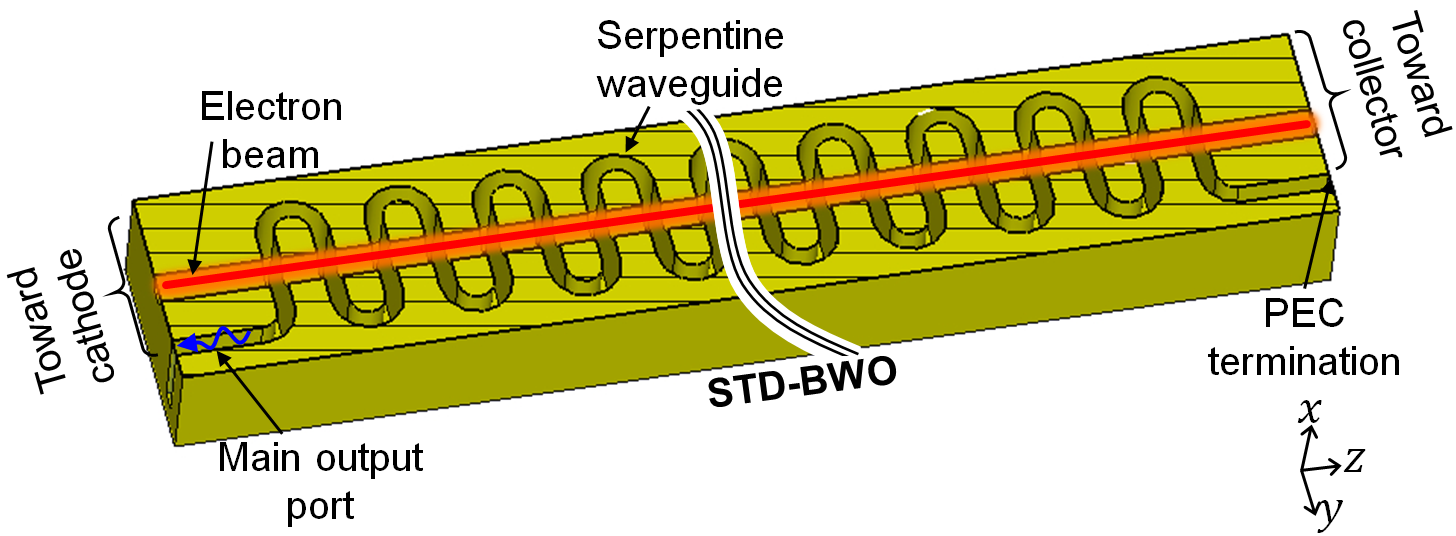}\label{Fig:SWS_No_DPE}} 
\par\end{centering}
\begin{centering}
\subfigure[]{\includegraphics[width=1\columnwidth]{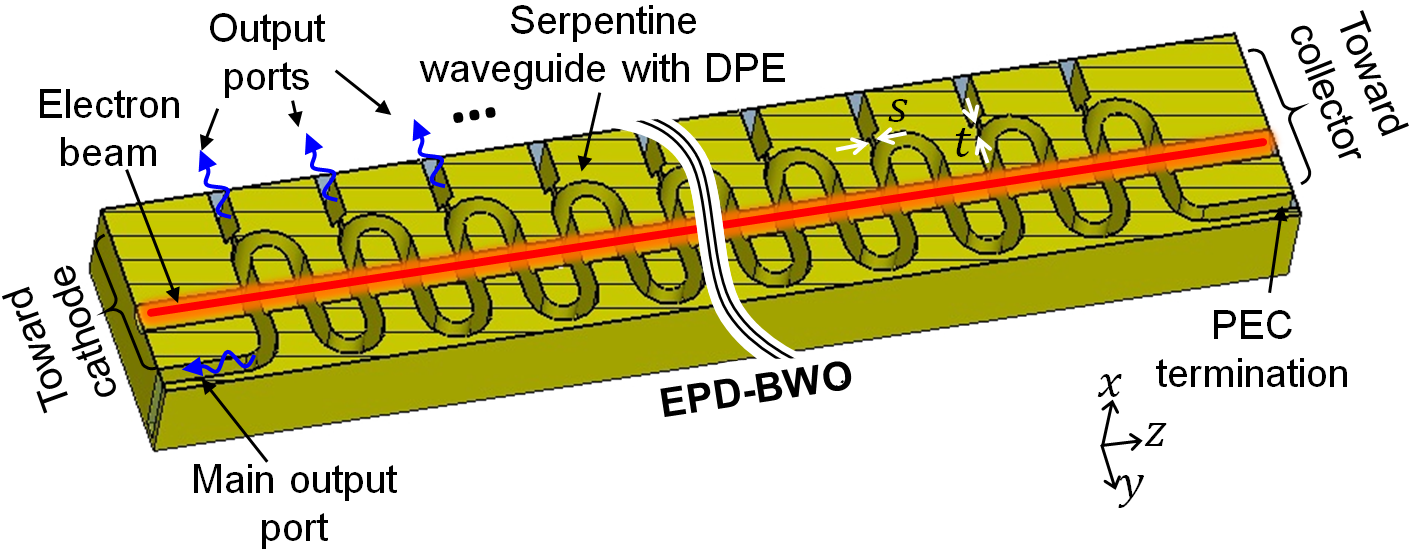}\label{Fig:SWS_DPE}} 
\par\end{centering}
\centering{}\caption{BWO with folded waveguide operating at millimeter waves: (a) standard
(STD)-BWO where the power is extracted from a waveguide end; (b) EPD-BWO
where the power is extracted in a distributed fashion to enable the
occurrence of the degenerate synchronization regime when working at
an EPD \cite{mealy2019backward,mealy2019exceptional,mealy2020exceptional}.
The distributed power is extracted by introducing a small slot in
each folded waveguide period that couples portion of the power in
the folded waveguide to the outgoing rectangular waveguides.}
\label{Fig:BWOs_All_SER}
\end{figure}

A simple model for the interaction between the e-beam and the EM wave
in vacuum tube devices was provided by Pierce in \cite{pierce1951waves}.
Augmenting the Pierce model to include SWSs with distributed loads,
we have shown in \cite{mealy2019backward,mealy2019exceptional,mealy2020exceptional}
that a second order EPD is found as a special degeneracy of the interactive
(hybrid) modes when DPE along the waveguide is added and when the
beam dc current is set to specific value $I_{0}=I_{0e}$, i.e., using
this e-beam dc current guarantees that two modes in the interactive
system are synchronized. The e-beam dc current $I_{0e}$ is the specific
value that guarantees the degeneracy of two modes, which can be set
to a desired value by properly designing the interactive SWS system.
The cold SWS with DPE has a complex propagation constant $\beta_{c}=\beta_{cr}+i\beta_{ci}$
around the synchronization point, where the imaginary part $\beta_{ci}$
accounts for the DPE along the SWS and the subscript $c$ denotes
circuit that is representing the cold SWS. We have shown in \cite{mealy2019backward,mealy2019exceptional,mealy2020exceptional}
that the EPD e-beam dc current has a proportionality $I_{0e}\propto\beta_{ci}^{2}$,
where the parameter $\beta_{ci}$ represents the DPE introduced in
the SWS and is determined by engineering the DPE from the SWS. The
fact that an EPD e-beam current $I_{0e}$ is found for any amount
of desired distributed power extraction implies that this so called
``degenerate synchronization'' regime is guaranteed for any desired
high power generation. Therefore, in principle, the synchronism is
maintained for any desired distributed power output, according to
the augmented Pierce-based model presented in \cite{mealy2019backward,mealy2019exceptional}.
Note that this trend is definitely not observed in standard BWOs where
interactive modes are non-degenerate and the load is at one end of
the SWS. Furthermore in a STD-BWO the starting (i.e., threshold) current
vanishes with increasing SWS length, whereas in an EPD-BWO it decreases
quadratically to a fixed, desired, value \cite{mealy2019backward,mealy2019exceptional,mealy2020exceptional},
that coincides with $I_{0e}\propto\beta_{ci}^{2}$, hence this value
is related to the amount of DPE. Therefore, the degenerate regime
enables a large transfer of power from the e-beam to the waveguide
EM mode as compared to a standard regime.

\begin{figure}
\begin{centering}
\subfigure[]{\includegraphics[width=0.45\columnwidth]{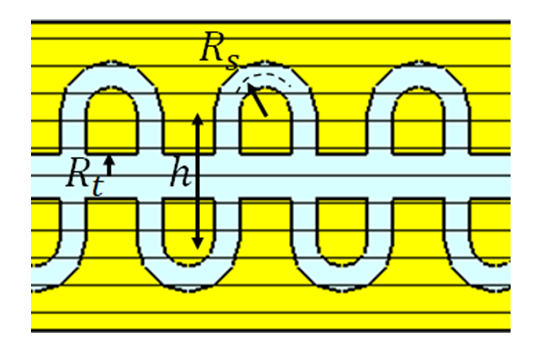}}
\subfigure[]{\includegraphics[width=0.45\columnwidth]{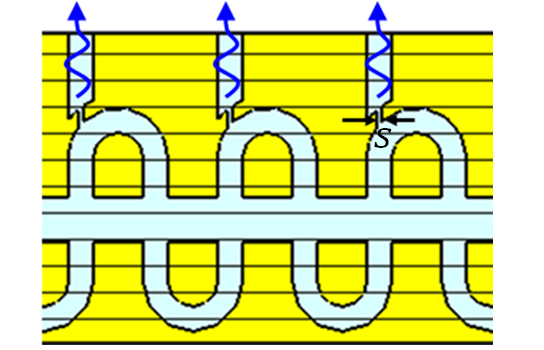}} 
\par\end{centering}
\begin{centering}
\subfigure[]{\includegraphics[width=0.98\columnwidth]{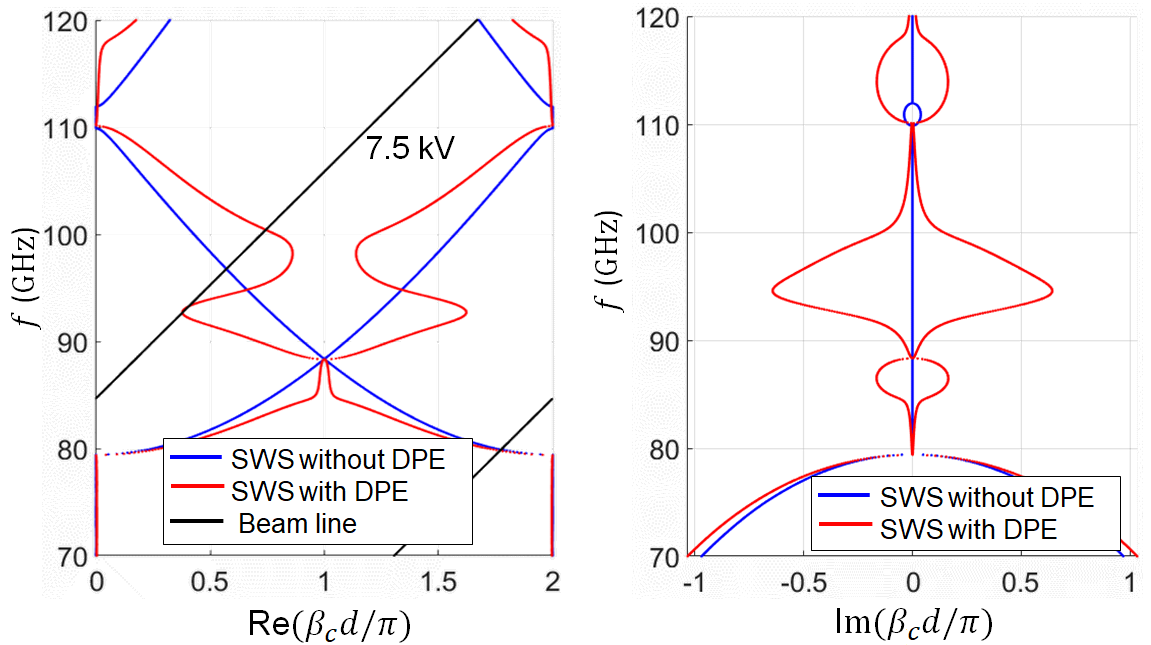}} 
\par\end{centering}
\centering{}\caption{Longitudinal cross-sections of a folded waveguide without (a) and
with DPE (b). (c) Dispersion of EM guided modes in the ``cold''
waveguide in (a) and (b), without (blue curve) and with (red curve)
DPE, respectively. The dispersion shows the real and imaginary parts
of the complex wavenumber. The non-zero imaginary part of the wavenumbers
of the EM mode with DPE (red line) shows that the waveguide in (b)
exhibits distributed power extraction. The imaginary part of the mode
without DPE (blue line) is almost zero.}
\label{Fig:Mix_2}
\end{figure}

Here, we show how to implement the EPD regime in a BWO that uses a
folded waveguide and operating at millimeter wave and THz frequency
through introducing DPE along the waveguide as illustrated in Fig.
\ref{Fig:SWS_DPE}. In particular, we consider a folded waveguide
made of copper with rectangular cross-section of dimension $a=1.9$
mm and $b=0.2$ mm. The folded waveguide has a bending radius of $R_{s}=0.3$
mm and the straight sections have length of $h=1$ mm. The beam tunnel
radius is $R_{t}=0.175$ mm, with a filling factor of about 58\%,
i.e., the e-beam has radius of $R_{b}=0.13$ mm. The DPE is conceived
in the waveguide by making a small rectangular slot of dimensions
$a\times s$, with $s=0.05$ mm, in the wide side of the rectangular
cross section, and length $t=0.1$ mm, in each period of the waveguide
(as shown in Fig. \ref{Fig:SWS_DPE}). The slots couples portion of
the main power in the SWS to an outgoing rectangular waveguide with
dimensions $a\times b$, as shown in Fig. \ref{Fig:Mix_2}(b).

Figure \ref{Fig:Mix_2}(c) shows a comparison between the dispersion
relation of EM modes in two ``cold'' SWSs: one used in a STD-BWO
in Fig. \ref{Fig:Mix_2}(a), and one used in the BWO with DPE in Fig.
\ref{Fig:Mix_2}(b). The dispersion curves show only the dominant
EM mode $\mathrm{TE_{10}}$ which exhibits an axial (longitudinal,
along the folded waveguide) electric field component able to interact
with the e-beam. The dispersion curves in Fig \ref{Fig:Mix_2}(c)
show that the cold SWS with DPE supports backward waves that have
a propagation constant with non-zero imaginary part $\beta_{ci}$
at the frequencies where the interaction with the e-beam may occur.
The imaginary part of a mode propagating in the folded waveguide without
DPE is almost equal to zero at the interaction points ($\omega/\beta_{cr}\approx u_{0}$,
where $u_{0}$ is the electrons average speed and $\omega$ is the
angular frequency). The complex wavenumber dispersion relation in
presence of DPE, shown in Fig \ref{Fig:Mix_2}(c), is obtained by
simulating a single unit-cell that is connected to two ports at its
beginning and end, while the power extraction waveguide is connected
to a matched port, and the beam tunnel ends are terminated by prefect
electric conductor (PEC) since the $\mathrm{TE_{10}}$ mode is below
cutoff of such a tunnel (we have checked that varying the beam tunnel
length and terminations does not affect the result). The Finite Element
Frequency Domain solver implemented in CST Studio Suite by DS SIMULIA
is used to calculate the two-port scattering parameters which are
converted into a two-port transfer matrix representing one unit-cell.
The complex Floquet-Bloch modes wavenumbers are then obtained by enforcing
periodic boundaries for the obtained transfer matrix, following the
method discussed in \cite{othman2016theory}.

\section{Particle-in-cell simulations of degenerate synchronous regime in
BWO}

We perform PIC simulations to assess the performance and features
of the proposed EPD-BWO using a folded waveguide and operating at
millimeter wave frequency. The PIC simulations uses a pencil e-beam
with dc voltage $V_{0}=7.5$ kV and an axial dc magnetic field of
2 T to confine the e-beam.

\subsection{Starting current}

\begin{figure}
\begin{centering}
\subfigure[]{\label{Fig:Output_No_Osc_No_DPE}\includegraphics[width=0.85\columnwidth]{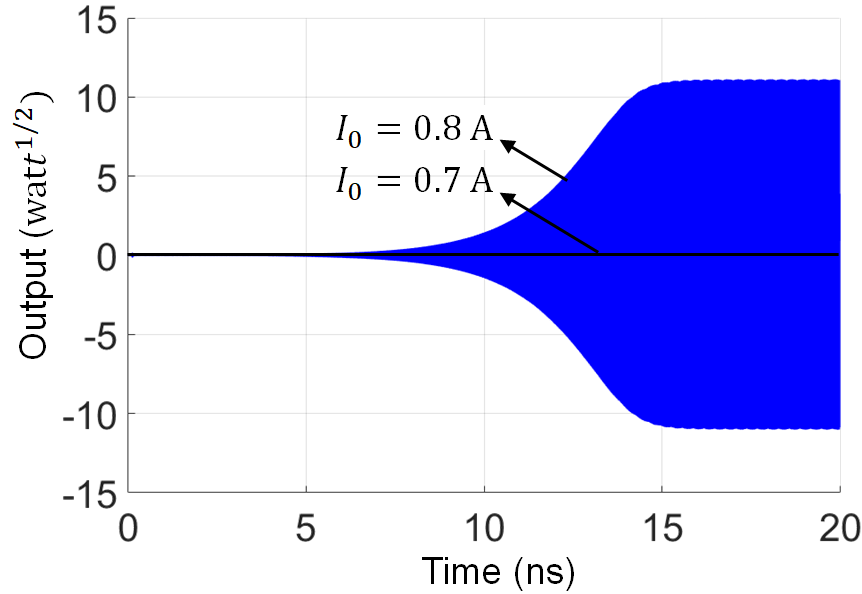}}
\par\end{centering}
\begin{centering}
\subfigure[]{\label{Fig:Output_Osc_No_DPE}\includegraphics[width=0.85\columnwidth]{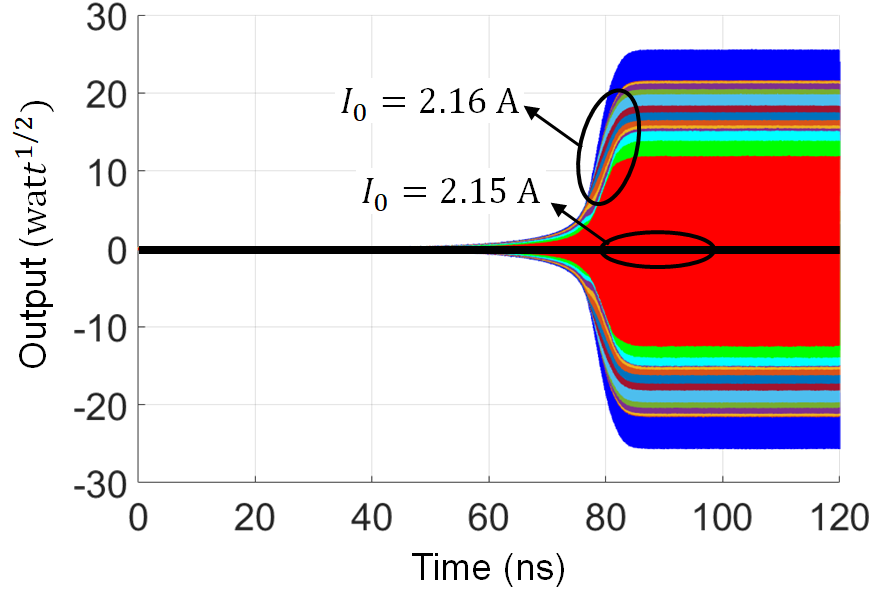}}
\par\end{centering}
\centering{}\caption{Output signals for: (a) STD-BWO and (b) EPD-BWO, both having 13 unit-cells,
when the e-beam dc current is just below and above the starting current
of oscillation ($I_{st}$) for each BWO case. The output signal for
the STD-BWO is extracted at the left-end port of the waveguide (a),
whereas for the EPD-BWO the signals are extracted from the 13 distributed
ports and from the left-end port, denoted by different colored curves
(b).}
\label{Fig:Output_Vs_Time}
\end{figure}

We start first by studying the starting (threshold) current of oscillation
to see how much increase in starting current we obtain with the EPD-BWO
with respect to a STD-BWO without DPE. Results show that the EPD-BWO
exhibits a much higher starting current of oscillation with respect
to a STD-BWO without DPE. Indeed the increase in starting current
is an advantage here because it enables to push up the saturation
power level of the BWO to much higher levels and therefore leads to
a high level of power extraction. The output power for a STD-BWO is
extracted from the main port at the left end of the waveguide, whereas
the output power of the EPD-BWO is extracted from the left waveguide
port in addition to all DPE waveguide ports as shown in Fig. \ref{Fig:SWS_DPE}.
Fig. \ref{Fig:Output_Vs_Time} shows the output ports signals, obtained
from PIC simulations, for both BWOs with 13 unit cells when the e-beam
current is just below and above the starting currents of each BWO
(the threshold currents have been found by repeating simulations with
varying e-beam dc current values with steps of 0.1 A for the STD-BWO
and 0.01 A for the EPD-BWO). A self-standing oscillation frequency
of 93.1 GHz is observed when the beam dc current $I_{0}$ is at or
larger than $0.8$ A for the STD-BWO case, whereas a self-standing
oscillation frequency of 88.2 GHz is observed when the beam dc current
$I_{0}$ is at or larger than $2.16$ A for the EPD-BWO case. We estimate
the starting current of the oscillation as the average of the two
observed values of the e-beam currents where oscillation starts to
occur and does not occur, respectively. Therefore,the starting currents
$I_{st}$ for STD-BWO and EPD-BWO are0.75 A and 2.15 A, respectively.
Considering larger and larger numbers of unit cells implies that $I_{st}\to0$
for the STD-BWO case, and $I_{st}\to I_{0e}$ for the EPD-BWO case
\cite{mealy2019exceptional,mealy2020exceptional}.

\begin{figure}
\begin{centering}
\includegraphics[width=0.99\columnwidth]{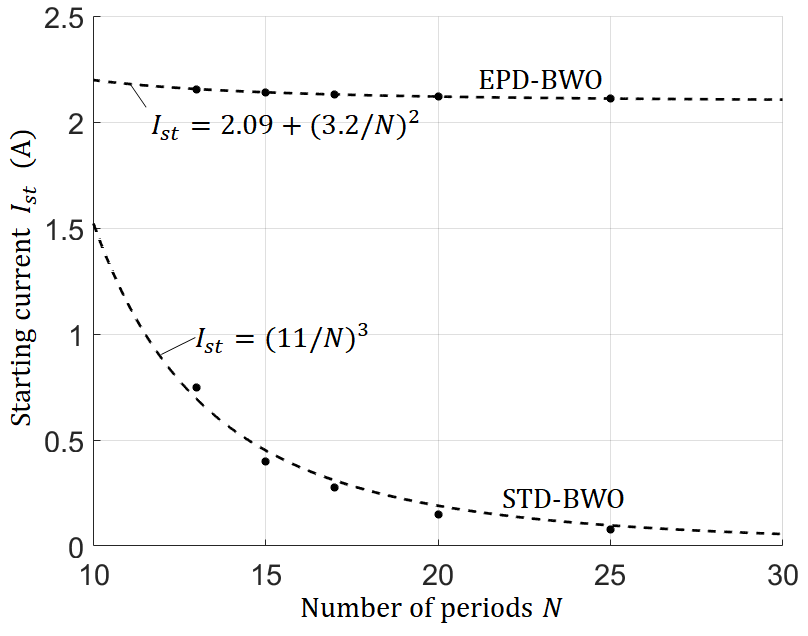}
\par\end{centering}
\centering{}\caption{Scaling of starting e-beam dc current for STD-BWO and EPD-BWO with
SWSs length (black dots). Dashed lines represent fitting curves. The
EPD-BWO shows a starting current trend that does not vanish for long
lengths, the quadratic decay is representative of a degeneracy condition.}
\label{Fig:Starting_Current}
\end{figure}

To assess the occurrence of an EPD, we verify the unique scaling trend
of the starting current in (\ref{Fig:Starting_Current}) by repeating
the previous study for different SWS lengths. Such scaling trends
for the starting current for both STD-BWO and EPD-BWO are shown using
black dots in Fig. \ref{Fig:Starting_Current} based on PIC simulation
results, varying the number of periods of the SWS. The dashed lines
represent fitting curves; the case of EPD-BWO shows very good agreement
with the fitting curve. The EPD-BWO is characterized by a starting
current that does not tend to zero as the SWS length increases, in
contrast to the starting current of a STD-BWO that vanishes for increasing
SWS length. The observed scaling of the starting current of a EPD-BWO
is a quadratic function of the inverse of the SWS length, which is
the same trend observed theoretically in \textcolor{black}{\cite{mealy2019exceptional}}
using an augmented Pierce model. Indeed, we have shown in \cite{mealy2019exceptional,mealy2020exceptional}
that the starting current of oscillation for EPD-BWO scales with the
SWS length as

\begin{equation}
I_{st}=I_{0e}+\left(\dfrac{\alpha}{N}\right)^{2},
\end{equation}
where $\alpha$ is a constant. From the fitting shown in Fig. \ref{Fig:Starting_Current}
we have found that $I_{st}=2.09+(3.2/N)^{2},$therefore, following
\textcolor{black}{\cite{mealy2019exceptional}}, the estimate of the
EPD current is $I_{0e}=I_{st}|_{N\to\infty}=2.09$ A. Later on in
the next section, we show that the use of a current close to this
value will lead to the degeneracy of two hybrid modes in the dispersion
of the hot structure, using data extracted from PIC simulations.

\subsection{Power performance: EPD-BWO compared to a STD-BWO}

\begin{figure}
\begin{centering}
\subfigure[]{\label{Fig:Pout_Vs_N}\includegraphics[width=0.995\columnwidth]{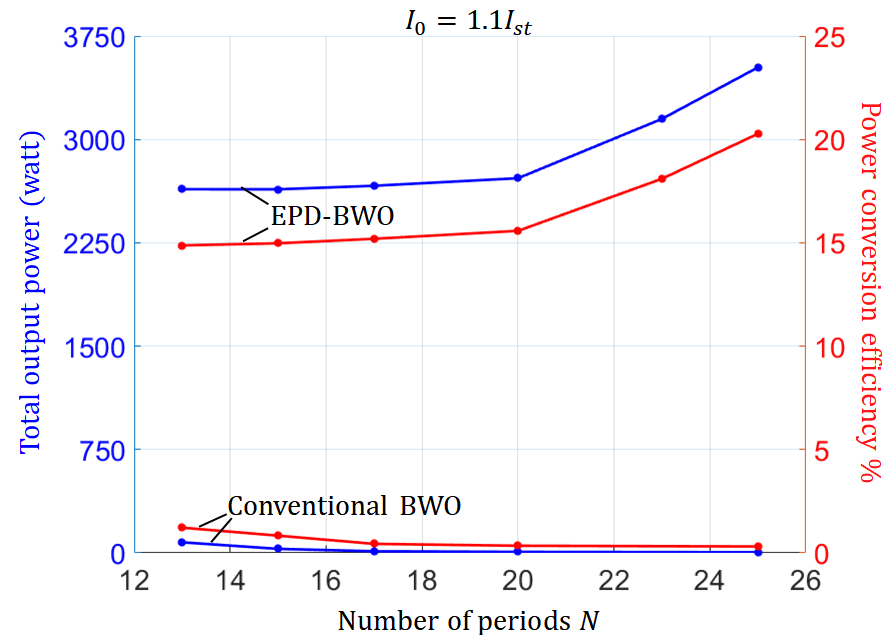}}
\par\end{centering}
\begin{centering}
\subfigure[]{\label{Fig:Pout_Vs_I}\includegraphics[width=0.995\columnwidth]{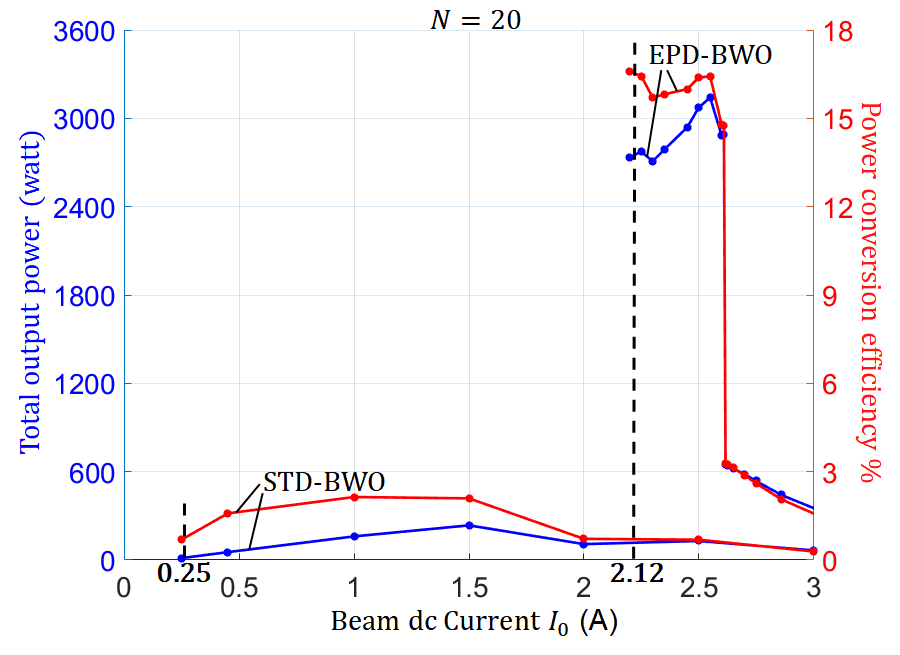}}
\par\end{centering}
\centering{}\caption{Comparison between the output power and power conversion efficiency
of a STD-BWO and an EPD-BWO, both based on the same folded waveguide
SWS, without and with DPE, respectively. In (a) we observe the power
trends when varying the number of unit cells of the folded waveguide
when the used beam dc current is 10\% higher than the starting currents
for the STD-BWO and EPD-BWO, and for each length. In (b) we observe
the power trends when varying the the beam dc current, assuming the
SWSs for the two BWOs are made of 20 unit cells. The figure shows
that EPD-BWO has much higher efficiency at much higher level of power
generation compared to the STD-BWO.}
\label{Fig:Power_Performance}
\end{figure}

We calculate the output power for STD-BWO and EPD-BWO when the e-beam
is 10\% above the starting beam current for each case, i.e., $I_{0}=1.1I_{st}$,
where the starting currents for both the EPD-BWO and STD-BWO are shown
in Fig. \ref{Fig:Starting_Current}. The output power $P_{out}$ for
the EPD-BWO case is calculated as the sum of the power delivered to
the distributed ports and the main port. We show in Fig. \ref{Fig:Pout_Vs_N}
the output power and power conversion efficiency, defined as $\eta=P_{out}/(V_{0}I_{0})$,
for both cases of EPD-BWO and STD-BWO when varying the number of periods
(i.e., unit cells) of the folded waveguide . The figure shows that
the EPD-BWO has always much higher output power level and power conversion
efficiency as compared to the STD-BWO. We observe from the figure
also that the output power and efficiency for the EPD-BWO increase
as we increase the number of folded waveguide periods, unlike for
the STD-BWO.

We then show in Fig. \ref{Fig:Pout_Vs_I} the output power and power
conversion efficiency for both cases of EPD-BWO and STD-BWO when changing
the beam dc current, keeping the number of period equal to $N=20$
for both kinds of BWOs. The figure shows that when the beam dc current
is exceeding the starting current for each of the EPD-BWO and STD-BWO,
the EPD-BWO is achieving much higher power conversion efficiency at
much higher level of power extraction. Note that we sweep the current
for a larger range for the STD-BWO case to be able reach the level
of currents that is used for EPD-BWO to be able to have a fair comparison.
Maximum output power is achieved for EPD-BWO case when the used current
is $I_{0}=2.6$ A which is about 22\% higher than the starting current
of oscillation and 24\% higher than the EPD current $I_{0e}$ estimated
from Fig. \ref{Fig:Starting_Current}. We expect that longer length
of the folded waveguide we use, the closer we are to the EPD and higher
efficiency and output power are obtained.

\section{Degenerate dispersion relation for the hot structure based on PIC
simulation }

\begin{figure}
\begin{centering}
\subfigure[]{\label{Fig:Disp_Setup}\includegraphics[width=0.9\columnwidth]{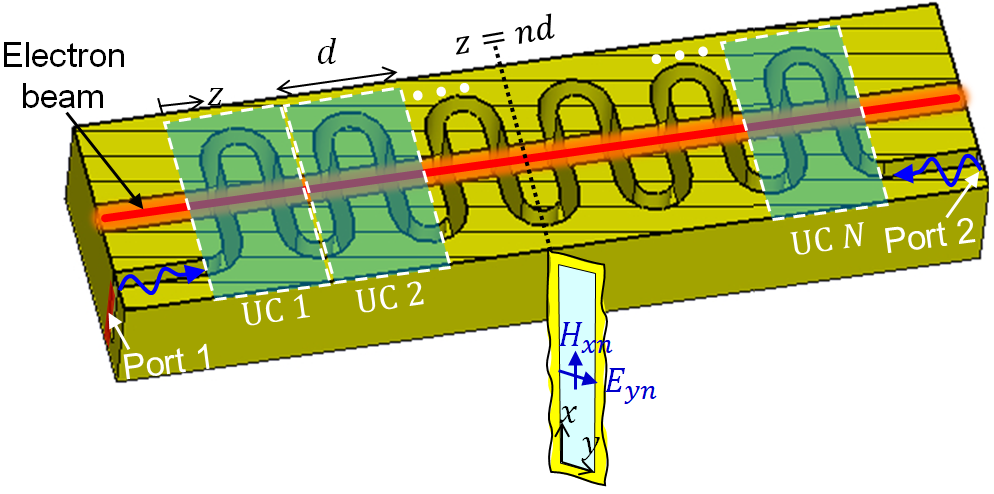}}
\par\end{centering}
\begin{centering}
\subfigure[]{\label{Fig:Disp_Setup_Model}\includegraphics[width=0.95\columnwidth]{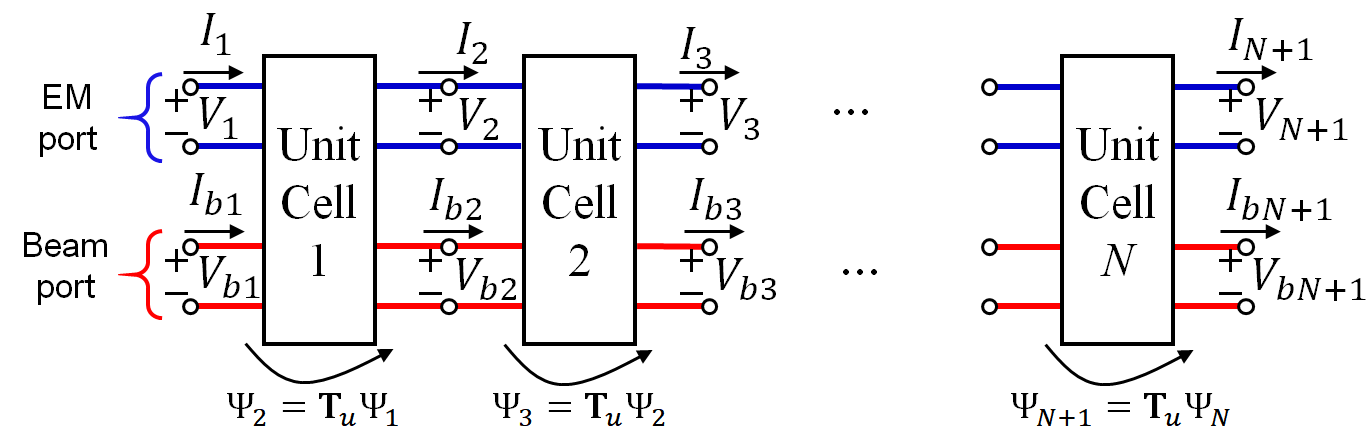}}
\par\end{centering}
\centering{}\caption{(a) Setup used to determine the complex-valued wavenumber versus frequency
dispersion relation of the hybrid modes in the hot SWS based on data
extracted from PIC simulations. (b) Circuit model showing that each
unit cell in the hot SWS is modeled as four-port network circuit with
\textit{equivalent} voltages and currents representing EM waves ($V_{n}$,
$I_{n}$) and space-charge waves ($V_{bn}$, $I_{bn}$) dynamics. }
\label{Fig:Starting_Current-1}
\end{figure}

The goal of this section is to show the degeneracy of the wavenumbers
of the modes of the interactive system (i.e, the hybrid modes) using
PIC simulations. Previously the degenerate dispersion has been shown
only using the approximate analytical method based on the Pierce model
in \cite{mealy2019exceptional}. Using the idealized analytical method
we also demonstrated the degeneracy of two eigenvectors at the EPD
\cite{mealy2019exceptional}. Here we adopt the general numerical
procedure described in \cite{mealy2020traveling} to estimate the
wavenumbers of the interactive (hybrid) modes, and show the hybrid
mode degeneracy using data extracted from PIC simulations. The advantage
is that PIC simulations predict the behavior of a realistic structure,
while the model in \cite{mealy2019exceptional} was just based on
Pierce theory. The procedure is based on exciting the structure from
both sides of the SWS by EM waves having monochromatic signal as illustrated
in Fig. \ref{Fig:Disp_Setup}, and then calculating the state vectors
that describe the EM field and the e-beam dynamics at discrete periodic
locations along the SWS. The time domain data extracted from PIC simulations
are transformed into phasors after reaching a steady state regime
as described in \cite{mealy2020traveling}. We then find the transfer
matrix associated to a unit-cell of the ``hot'' SWS that best relates
the calculated state vectors at both ends of each unit cell. Once
the estimate of the unit-cell transfer matrix is obtained, we find
the hybrid eigenmodes in a hot SWS using Floquet theory. We provide
details about the steps we used to generate the dispersion relation
for the hot SWS in Appendix A.

\begin{figure}
\begin{centering}
\subfigure[]{\includegraphics[width=1\columnwidth]{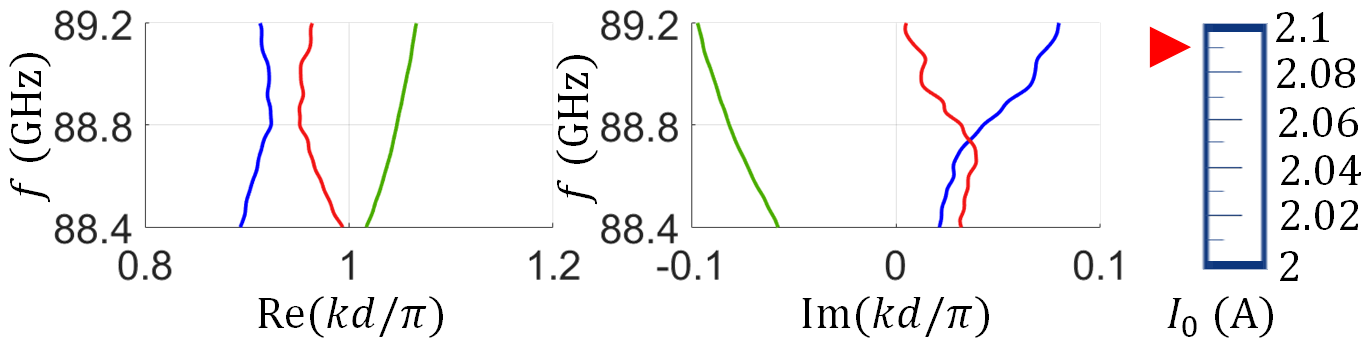}}
\par\end{centering}
\begin{centering}
\subfigure[]{\includegraphics[width=1\columnwidth]{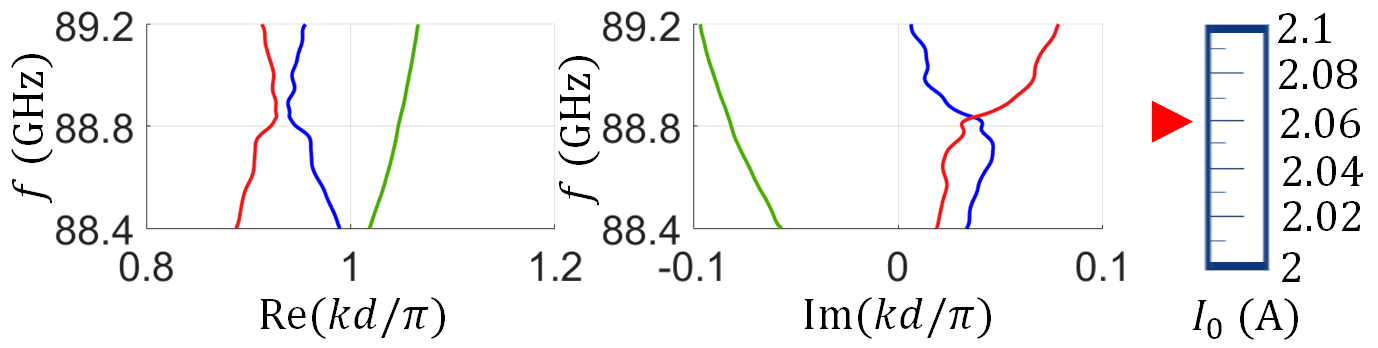}}
\par\end{centering}
\begin{centering}
\subfigure[]{\includegraphics[width=1\columnwidth]{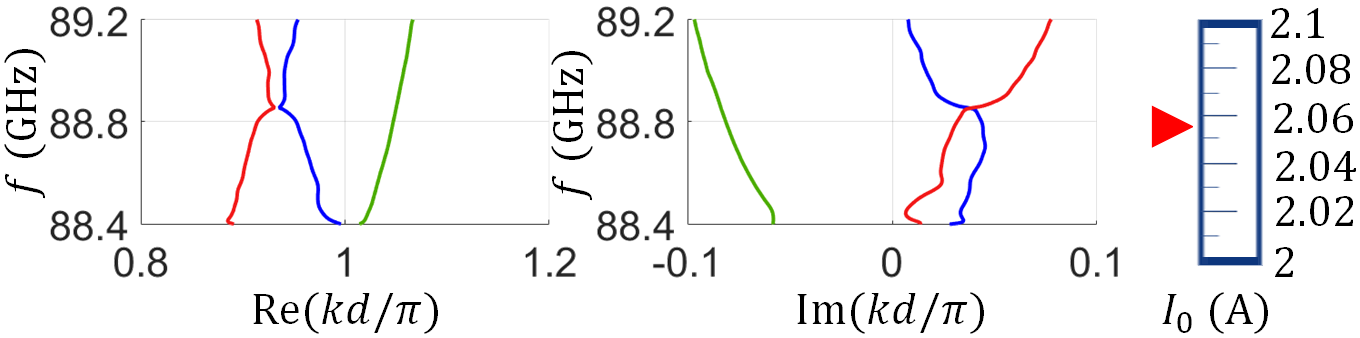}}
\par\end{centering}
\begin{centering}
\subfigure[]{\includegraphics[width=1\columnwidth]{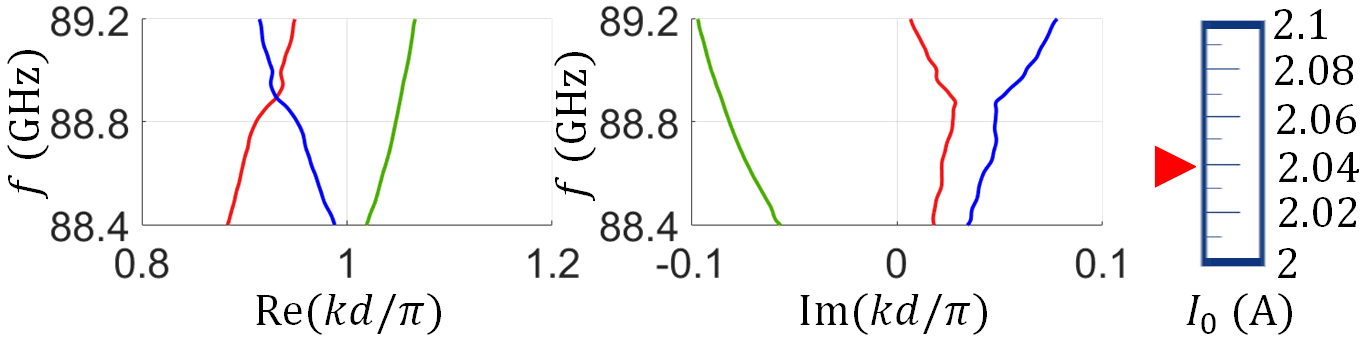}}
\par\end{centering}
\begin{centering}
\subfigure[]{\includegraphics[width=1\columnwidth]{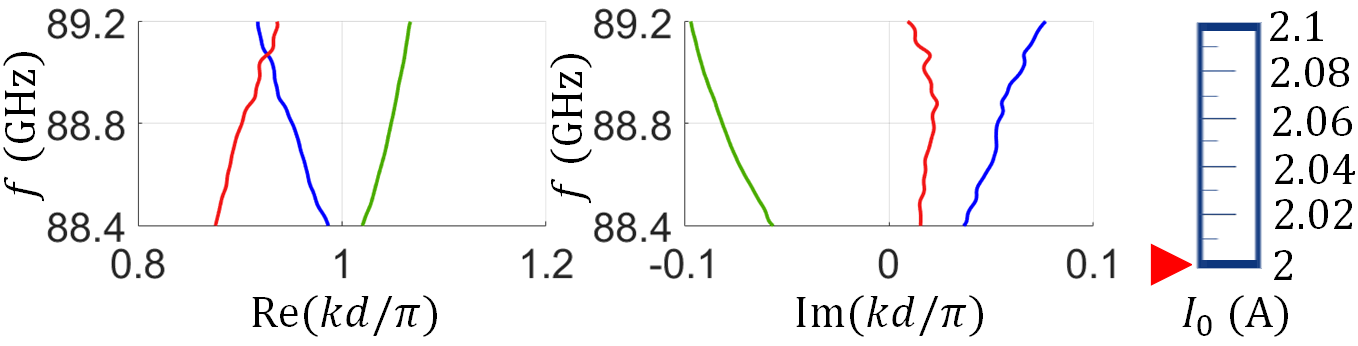}}
\par\end{centering}
\centering{}\caption{Dispersion of complex-valued wavenumbers of three hybrid modes showing
the wavenumber degeneracy at about $I_{0}=2.055$A. Wavenumbers of
the three modes with positive real part are retrieved from data obtained
by PIC simulations relative to hot SWSs with 11 unit-cells, when using
different e-beam dc current: (a)$I_{0}=2.09$A, (b)$I_{0}=2.06$A,
(c)$I_{0}=2.055$A, (d)$I_{0}=2.04$ and (e)$I_{0}=2.00$A. All the
considered beam dc currents to generate the results are lower than
the starting current of oscillation which is estimated to be 2.175
A when using 11 unit cells. The plots show a modal degeneracy when
the the beam dc current is about $I_{0}=2.055$A, which is very close
to the EPD current value of $I_{0e}=2.09$ estimated using the fitting
in Fig. \ref{Fig:Starting_Current}, and the operating frequency at
which the degeneracy is observed is about $f=88.8$ GHz, which is
close to the oscillation frequency.}
\label{Fig:DISP_Relation}
\end{figure}

The wavenumber-frequency dispersion describing the complex-valued
wavenumber of the hybrid eigenmodes in the hot SWS is determined by
running multiple PIC simulations for a SWS made of 11 unit-cells,
at different frequencies and at different beam dc currents, and then
determining the transfer matrix of the unit-cell at each frequency
and current combination using Eq. (\ref{eq:Best_Fit}). We sweep the
e-beam dc current around the expected value of EPD beam current $I_{0e}$,
which is the value of e-beam current pertaining to the infinitely
long SWS according to the fitting shown in Fig. \ref{Fig:Starting_Current};
the use a current that is close to this current value should guarantee
the coalescence of two interactive beam-EM modes. It is important
to mention that the used beam currents to generate the results in
this section are below the starting current of oscillation which is
estimated to be 2.175 A as shown in Fig. \ref{Fig:Starting_Current},
in order to avoid strong saturation regimes proper of oscillators'
dynamics. Therefore, by neglecting nonlinearities, one models each
unit-cell of the hot SWS using a transfer matrix as discussed previously
and in Appendix A. Since the transfer matrix has dimension 4x4, there
are four eigenvalues, i.e., for complex-valued wavenumbers associated
to the four hybrid modes supported by the model shown in Fig. \ref{Fig:Disp_Setup_Model}.
We plot only the three modes that have a wavenumber with a positive
real part. The dispersion diagram of such three interactive modes
in the hot EM-electron beam system is shown in Fig. \ref{Fig:DISP_Relation}
at different e-beam dc current. The figure shows a degeneracy of two
hybrid modes (the red and blue curves) at a frequency near $f=88.8$
GHz which is very close to the oscillation frequency when the used
e-beam dc current is around $I_{0}=2.055$A which is close to (and
slightly lower) the value of EPD beam current value of $I_{0e}=I_{st}|_{N\to\infty}=2.09$
obtained from the fitting shown in Fig. \ref{Fig:Starting_Current},
calculated as a limit for an infinitely long SWS. All the e-beam currents
$I_{0}$ considered show that two wavenumbers are either close to
the degeneracy or degenerate. There is only a very small discrepancy
between the estimate of EPD current $I_{0e}$ obtained from the fitting
of the starting currents varying length and from observing the degeneracy
of two hybrid modes shown in Fig. \ref{Fig:DISP_Relation}. Such small
discrepancy could be attributed to the use of finite precision in
calculating the starting current of oscillation, or to the approximations
implied in the retrieval method used to obtain the dispersion of hot
structure, like possible nonlinearites that are not accounted for
in our retrieval model.

\section{\textcolor{black}{Conclusion}}

We have conceived a degenerate synchronization regime to increase
the output power and power conversion efficiency of BWOs operating
at millimeter wave and THz frequencies. The degenerate synchronization
regime is achieved through altering the folded waveguide by adding
periodic power extraction ports. This allows the interactive system
to work at an EPD which implies a maintained synchronism designed
for any desired level of power extraction. PIC simulation results
shows that a millimeter wave BWO operating at degenerate synchronization
regime is capable of generating output power exceeding 3 kwatts with
conversion efficiency of exceeding 20\% at a frequency of 88.5 GHz.
The unique quadratic starting current scaling law with waveguide interaction
lengths observed from PIC simulations demonstrates the EPD-based synchronization
phenomenon, compared to that in a STD-BWO that has a starting current
law that vanishes cubically. The complex-valued wavenumber degeneracy
is confirmed by elaborating data extracted from PIC simulations which
implies the existence of EPD in the interactive system. This new degenerate
synchronization regime may pave the way to the realization of very
high power sources at millimeter and submillimeter waves, with high
power conversion efficiency.

\section{Acknowledgment}

This material is based upon work supported by the Air Force Office
of Scientific Research award number FA9550-18-1-0355. The authors
are thankful to DS SIMULIA for providing CST Studio Suite that was
instrumental in this study.

\appendices{}

\section{Method used to find the dispersion of hybrid hot modes using data
from PIC simulations}

We provide the basic steps of the retrieval method used to generate
the dispersion relations for hot SWSs based on time domain data obtained
from PIC simulations. The reader is addressed to \cite{mealy2020traveling}
for more details.

We define a state vector that describes the EM and space-charge waves,
defined at discrete periodic locations $z=z_{n}=nd$ as

\begin{equation}
\mathbf{\Psi}_{n}=[\begin{array}{cccc}
V_{n}, & I_{n}, & V_{bn}, & I_{bn}\end{array}]^{T},\label{eq:FD-StateVect-1}
\end{equation}
where $V_{n}$ and $I_{n}$ are equivalent voltages and currents representing
the EM mode in the SWS, and $V_{bn}$ and $I_{bn}$ are equivalent
voltages and currents representing the charge wave modulating the
e-beam.

We assume that the EM field in the folded rectangular waveguide is
dominated by a $\mathrm{TE_{10}}$. Thus the (locally-) transverse
field at the rectangular cross section is in the form of

\begin{equation}
\begin{array}{c}
\mathbf{E}_{tn}(x,y)=E_{yn}\sin\left(\dfrac{\pi x}{a}\right)\mathbf{a}_{y},\\
\mathbf{H}_{tn}(x,y)=H_{xn}\sin\left(\dfrac{\pi x}{a}\right)\mathbf{a}_{x},
\end{array}\label{eq:Fields}
\end{equation}

The coordinates used here do not described the waveguide folding,
we just use them to describe the field at cross sections of the rectangular
waveguide at discrete locations $z=z_{n}$. We use an equivalent transmission
line model that has voltage and current $V_{n}$ and $I_{n}$ representing
the EM field in the periodic SWS. The voltage and current are calculated
as the projection of the transverse electric and magnetic fields in
the SWS on the electric and magnetic transverse eigenvector basis
proper of a $\mathrm{TE_{10}}$ mode as explained in \cite{marcuvitz1951waveguide,collin1990field,felsen1994radiation}

\begin{equation}
\begin{array}{c}
V_{n}=\langle\mathbf{E}_{tn}(x,y),\mathbf{e}(x,y)\rangle,\\
I_{n}=\langle\mathbf{H}_{tn}(x,y),\mathbf{h}(x,y)\rangle,
\end{array}
\end{equation}
where $\langle\mathbf{A},\mathbf{B}\rangle=\int_{0}^{a}\int_{0}^{b}\mathbf{A}\cdot\mathbf{B}^{*}dxdy$
represents the field projection onto the chosen transverse eigenvector
basis. The electric and magnetic field transverse basis are calculated
as $\mathbf{h}(x,y)=-\nabla_{t}\Phi(x,y)$ and $\mathbf{e}(x,y)=\mathbf{u}_{z}\times\mathbf{e}(x,y)$
\cite{marcuvitz1951waveguide,collin1990field,felsen1994radiation},
and $\Phi$ is the scalar basis function for the $\mathrm{TE_{10}}$
mode and it is found as $\Phi(x,y)=\Phi_{0}\sin\left(\pi x/a\right)$
and $\Phi_{0}$ is determined by enforcing normality of the electric
field basis as $\langle\mathbf{e},\mathbf{e}\rangle=1$ \cite{marcuvitz1951waveguide,collin1990field,felsen1994radiation}.
Therefore the electric and magnetic transverse eigenvectors are 
\begin{equation}
\begin{array}{c}
\mathbf{e}(x,y)=\sqrt{\dfrac{2}{ab}}\sin\left(\dfrac{\pi x}{a}\right)\mathbf{a}_{y},\\
\mathbf{h}(x,y)=\sqrt{\dfrac{2}{ab}}\sin\left(\dfrac{\pi x}{a}\right)\mathbf{a}_{x}.
\end{array}\label{eq:Basis}
\end{equation}
The the voltage and the current of the equivalent transmission line
are calculated from the transverse electric and magnetic fields at
the center of the rectangular waveguide cross-section by projecting
fields in (\ref{eq:Fields}) on those in (\ref{eq:Basis})

\begin{equation}
\begin{array}{c}
V_{n}=\sqrt{\dfrac{ab}{2}}E_{yn},\\
I_{n}=\sqrt{\dfrac{ab}{2}}H_{xn}.
\end{array}
\end{equation}

Although a PIC solver calculates the speeds of the discrete large
number of charged particles, we represent the longitudinal speed of
all e-beam charges as one dimensional, i.e., as a scalar function.
The beam total equivalent kinetic voltage at the entrance of the $n^{th}$
unit-cell is defined in time domain as $v_{bn}^{tot}(t)=\sqrt{2\eta u_{bn}^{tot}(t)},$
where $u_{bn}^{tot}(t)$ is the average of the speed of the charges
at each \textit{z}-cross section (see \cite{mealy2020traveling} for
more details). The ac modulation is then calculated as $v_{bn}(t)=v_{bn}^{tot}(t)-V_{0}$,
which is then converted to the phasor domain $V_{bn}$ to construct
the system's state vector. The term $I_{bn}$ is calculated as the
phasor domain transformation of the ac part of the current of the
e-beam, at the discrete locations $z=z_{n}=nd$ (see \cite{mealy2020traveling}
for more details).

In the phasor-domain, we model each unit cell of the interacting SWS
as a 4-port network circuit as shown in Fig. \ref{Fig:Disp_Setup_Model},
and we need to determine its associated transfer matrix. Under the
assumption of small signal modulation of the beam\textquoteright s
electron velocity and charge density, all the 4-port networks modeling
the interaction between the EM and the charge wave in each unit-cell
of the hot SWS are assumed to be identical. Therefore, the single
$4\times4$ transfer matrix $\mathbf{T}_{\mathit{u}}$ of the interaction
unit-cell should satisfy

\begin{equation}
\begin{array}{c}
\mathbf{\Psi}_{2}=\mathbf{T}_{\mathit{u}}\mathbf{\Psi}_{1},\ \ \ \ \ \ (\ref{eq:Tmatrix-StateVector-1}.1)\\
\mathbf{\Psi}_{3}=\mathbf{T}_{\mathit{u}}\mathbf{\Psi}_{2},\ \ \ \ \ \ (\ref{eq:Tmatrix-StateVector-1}.2)\\
\vdots\ \ \ \ \ \ \ \ \ \ \ \ \ \ \ \\
\mathbf{\Psi}_{N+1}=\mathbf{T}_{\mathit{u}}\mathbf{\Psi}_{N},\ \ \ \ \ \ (\ref{eq:Tmatrix-StateVector-1}.N)
\end{array}\label{eq:Tmatrix-StateVector-1}
\end{equation}
where $\mathbf{\Psi}_{n+1}$ and $\mathbf{\Psi}_{n}$ are the input
and output state vectors of the $n^{th}$ unit-cell, respectively,
with \textit{n} = 1, 2,..\textit{ N}. Since the state vectors are
calculated using data from PIC simulations, the relations in (\ref{eq:Tmatrix-StateVector-1})
represent $4N$ linear equations in $16$ unknowns, which are the
unknown elements of the transfer matrix $\mathbf{T}_{u}$. The system
in (\ref{eq:Tmatrix-StateVector-1}) is mathematically referred to
as overdetermined because the number of linear equations ($4N$ equations)
is greater than the number of unknowns (16 unknowns). We rewrite (\ref{eq:Tmatrix-StateVector-1})
in matrix form as $\left[\mathbf{W}_{2}\right]_{4\times N}=\left[\mathbf{T}_{\mathit{u}}\right]_{4\times4}\left[\mathbf{W}_{1}\right]_{4\times N}.$
The column of the matrices $\mathbf{W}_{1}$ and $\mathbf{W}_{2}$
are the state vectors at input and output, respectively, of each unit-cell
and they are written in the form $\mathbf{W}_{1}=\left[\begin{array}{cccc}
\mathbf{\Psi}_{1}, & \mathbf{\Psi}_{2}, & \ldots & \mathbf{\Psi}_{N}\end{array}\right]$ and $\mathbf{W}_{2}=\left[\begin{array}{cccc}
\mathbf{\Psi}_{2}, & \mathbf{\Psi}_{3}, & \ldots & \mathbf{\Psi}_{N+1}\end{array}\right]$ . An approximate solution that best satisfies all the given equations
in Eq. (\ref{eq:Tmatrix-StateVector-1}), i.e., it minimizes the sums
of the squared residuals, $\sum_{n}\left|\left|\mathbf{\Psi}_{n+1}-\mathbf{T}_{\mathit{u}}\mathbf{\Psi}_{n}\right|\right|^{2}$,
is determined similarly as in \cite{forsythe1977computer,williams1990overdetermined,anton2013elementary}
and is given by

\begin{equation}
\mathbf{T}_{\mathit{u}}=\left(\left[\mathbf{W}_{2}\right]_{4\times N}\left[\mathbf{W}_{1}\right]_{4\times N}^{T}\right)\left(\left[\mathbf{W}_{1}\right]_{4\times N}\left[\mathbf{W}_{1}\right]_{4\times N}^{T}\right)^{-1}.\label{eq:Best_Fit}
\end{equation}

The hybrid e-beam-EM eigenmodes are determined by assuming a state
vector has the form of $\mathbf{\Psi_{\mathrm{\mathit{n}}}}\propto e^{-jknd}$,
where $k$ is the complex Bloch wavenumber that has to be determined
and $d$ is the SWS period. Inserting the assumed sate vector \textit{z}-dependency
in (\ref{eq:Best_Fit}), we define the eigenvalue problem and we find
the four Floquet-Bloch mode wavenumbers, determined as 

\begin{equation}
e^{-jkd}=\mathrm{eig}(\mathbf{T}_{\mathit{u}}).\label{eq:Eigenvalue}
\end{equation}

The above equation lead to four Floquet-Bloch modes wavenumbers $k_{m}$,
where $m=1,2,3,4$, with harmonics $k_{m,}+2\pi q/d$, where $q$
is an integer that defines the spatial harmonic index.

\bibliographystyle{ieeetr}
\bibliography{myref}

\end{document}